\documentclass{iau}

\usepackage{amsmath}
\usepackage{graphicx}
\usepackage{multirow}
\usepackage{titlesec}
\titleformat{\section}{\large\bfseries}{\thesection}{1em}{}

\begin{document}

\lefttitle{Lekshmi et al.}
\righttitle{Temporal Variation of Solar Equatorial Rossby Modes}

\jnlPage{1}{7}
\jnlDoiYr{2021}
\doival{10.1017/xxxxx}

\aopheadtitle{Proceedings IAU Symposium}
\editors{A. V. Getling \&  L. L. Kitchatinov, eds.}

\title{Temporal Variation of Solar Equatorial Rossby Modes with Azimuthal Orders $6\leq m \leq 10$}

\author{B. Lekshmi$^{1}$, Laurent Gizon$^{1,2}$, Kiran Jain$^{3}$, Zhi-Chao Liang$^{1}$,  Jordan Philidet$^{1}$}

\affiliation{$^1$Max-Planck-Institut f\"ur Sonnensystemforschung,  37077 G\"ottingen, Germany \\ email: {\tt lekshmib@mps.mpg.de} \\
$^2$Institut f\"ur Astrophysik, Georg-August-Universität G\"ottingen, 37077 G\"ottingen, Germany \\
$^3$National Solar Observatory, Boulder, CO 80303, USA}

\begin{abstract}

We use nearly two decades  of helioseismic data obtained from the GONG (2002--2020) and HMI (2010--2020) ring-diagram pipelines to examine the temporal variations of the properties of individual  equatorial Rossby modes with azimuthal orders in the range $6 \le m \le 10$. We find that  the mode parameters obtained from GONG and HMI are consistent during the data overlapping period of 2010--2020.  The power and the frequency of each mode exhibit significant temporal variations over the full observing period. Using the GONG data during solar cycles 23 and 24, we find that the mode power averaged over $6 \le m \le 10$ shows a positive correlation with the sunspot number ($0.42$), while the averaged frequency shift  is anti-correlated ($-0.91$). The anti-correlation between the average mode power and frequency shift is $-0.44$.
\end{abstract}

\maketitle

\section*{Introduction}
The solar equatorial Rossby modes  were first observed using helioseismic observations from the Helioseismic and Magnetic Imager (HMI) onboard the Solar Dynamics Observatory  \citep{Loeptien2018, Liang2019, Gizon2021} and the  Global Oscillation Network Group (GONG) \citep{Hanson2020}. In the Carrington frame, these modes closely follow the standard dispersion relation for classical sectoral modes, $\nu \approx -2\nu_0/(m+1)$ where $\nu_0$ is the equatorial rotational frequency, $m$ is the azimuthal order, and the negative sign indicates retrograde propagation. 

\cite{Waidele2023}, using velocities obtained from time-distance and ring-diagram analysis of HMI data, measured solar-cycle variations in the mode amplitudes. They found that the average mode amplitude over $3 \leq m \leq 16$ positively correlates with the sunspot number. They also reported that the mode frequencies vary over the solar cycle, with more negative frequencies during solar maximum. 

The present study focuses on the temporal variation of the power and the frequency of  individual equatorial Rossby modes with  $m=6, 7, \cdots, 10$. In addition, we consider both the HMI and the GONG datasets.

\section*{ Data and Analysis}
We use the horizontal flows in the co-latitudinal direction ($u_\theta$) from the ring diagram (RD) pipelines of GONG \citep{Corbard2003} and HMI \citep{Bogart2011a}, covering the periods January 2002 to June 2020 and January 2011 to June 2020, respectively. The data extend over $\pm52.5^{\circ}$ in latitude and central meridian distance for GONG and $\pm67.5^{\circ}$ for HMI. We remove the zero and annual frequencies in the flows following the method of \cite{Proxauf2020}. An additional one-day periodicity aliased as $8.33$~days in the GONG measurements is also removed. We obtain the horizontal flows at a depth of $\sim 2$~Mm in the Carrington frame. We denote the component of velocity in the colatitudinal direction 
symmetrized across the equator by $u^{+}_\theta(\theta,\phi,t)$, where $\theta$ is the colatitude, $\phi$  the Carrington longitude, and $t$ time.

The GONG and HMI time series are divided into overlapping segments of duration $T=4$~years, shifted by multiples of  six months. For GONG, the central times of the segments are given by $t_n= 1.1.2004 + n \times 6$~months with $n=0,1,\cdots, 31$. For HMI,  the central times are $t_n= 1.1.2013 + n \times 6$~months with $n=0,1,\cdots, 12$.

The corresponding power spectra are  given by,
\begin{equation}
    P_m(\nu,t_n) = \frac{1}{N_{\theta}}\sum_{|\theta-90^\circ| \leq 30^\circ}  \left| \sum_{\phi,t^{\prime}} \textrm{rect}\left( \frac{t'-t}{T} \right)  u_{\theta}^{+}(\theta,\phi,t^{\prime})e^{\textrm{i}(m\phi-2\pi\nu t^{\prime})}\right|^2 ,
    \label{eq:FFT}
\end{equation}
where `rect' is a symmetric rectangular window function of width $1$, $N_\theta=3$ is the number of independent latitude bins over which the power is averaged. %and convert the units to $\rm m^2s^{-2}nHz^{-1}$. 

We also compute a reference power spectrum for each $m$ obtained by averaging over the data segments. The peak of power associated with the Rossby mode in each average power spectrum is fitted with a Lorentzian profile,
\begin{equation}
   % \overline{F}_\theta^+(m,\nu) = 
    L_m(\nu)=
    \overline{A}_m \left[ 1 + \left(\frac{\nu-\overline{\nu}_m}{\overline{\Gamma}_m/2}\right)^{2} \right]^{-1} + \overline{B}_m ,
    \label{eq:Lorentzian}
\end{equation}
where $\overline{A}_m$, $\overline{\nu}_m$, $\overline{\Gamma}_m$ and $\overline{B}_m$ are the mode amplitude,  frequency, full width at half maximum, and  background power respectively. 
The probability density function of the power spectrum at fixed frequency is approximated by a $\chi^2$ with $2\times N \times  N_\theta$ degrees of freedom, where $N$ is the number of independent time segments and the factor $N_\theta$ is for the 
number of independent  latitude bins. 
\begin{figure*}[h]
    \centering
    \includegraphics[width=\textwidth]{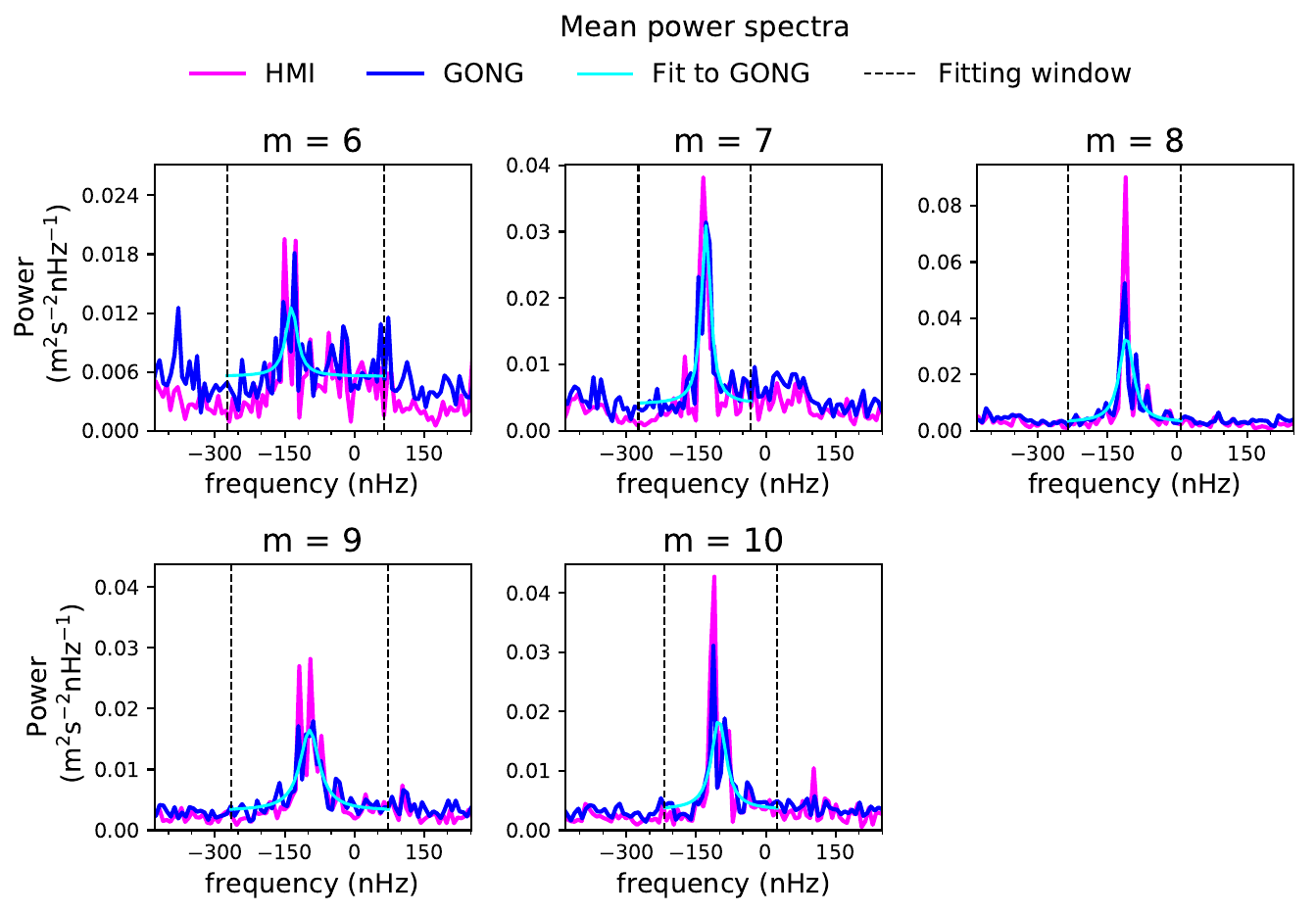}
    \caption{Power spectra of the equatorial Rossby modes for each $m$. The power spectra are averaged over all segments of duration $T=4$~years (frequency resolution is approximately 8~nHz). The dark blue curves are for GONG, the magenta curves for HMI. The Lorentzian fits to the GONG power spectra are overplotted in cyan. The fitting window is bounded by the two vertical dashed lines.}
    \label{fig:mean_spectra}
\end{figure*}

The GONG and HMI power spectra, together with the fit to the GONG power spectra, are shown in Fig.~\ref{fig:mean_spectra}. The fitting frequency range is chosen for each mode separately (see vertical dashed lines in Fig.~\ref{fig:mean_spectra}). 
We find that the mode parameters extracted from the GONG and HMI power spectra are consistent.

\begin{figure*}[h]
    \centering
    \includegraphics[width=\textwidth]{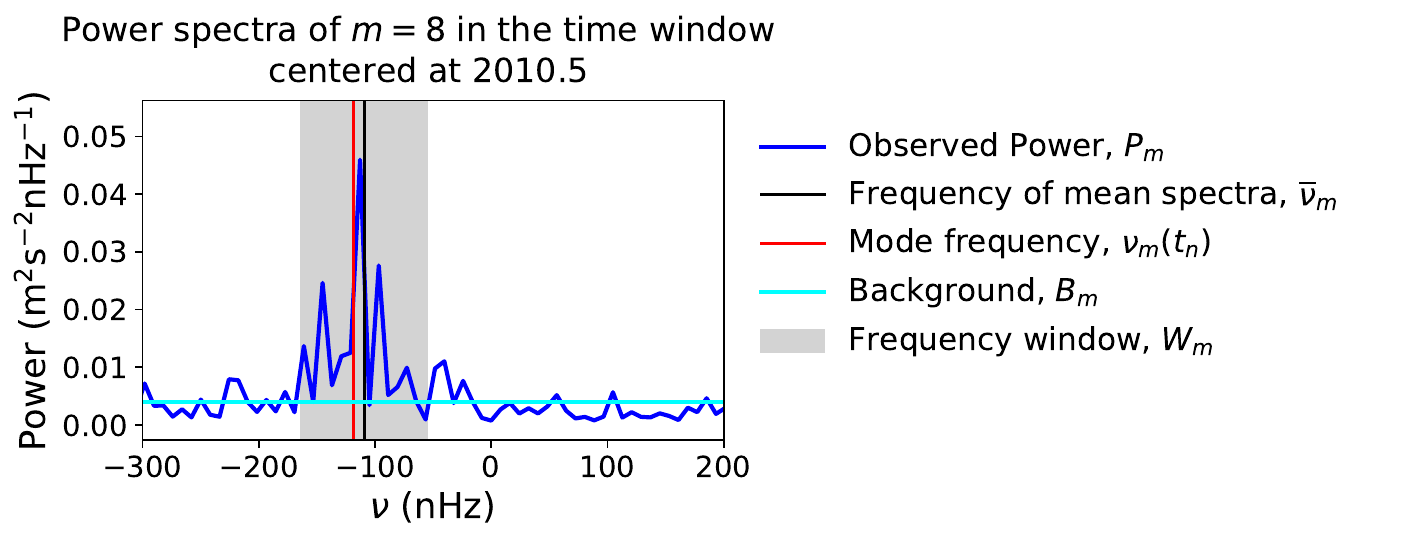}
    \caption{The GONG power spectrum of the $m=8$ mode in the four year time window centered at $t_n=2010.5$. The average mode frequency $\overline{\nu}_m$ for the entire data set is given by the vertical black line. The  mode frequency $\nu_m(t_n)$ obtained from Eq.~\ref{fig:totpow_freq} for this particular time stamp is marked  by the vertical  red line. The cyan horizontal line shows the estimated background $B_m(t_n)$ .}
    \label{fig:m8_spectra_tstamp}
\end{figure*}

To study the temporal variation of the mode parameters, we  compute the mode power ($E_m$) and the mode frequency ($\nu_m$) of the modes in each time segment as follows:
\begin{eqnarray}
    E_m (t_n) &=& \sum_{\nu\in W_m} \left[ P_m(\nu,t_n) - B_m(t_n) \right] \Delta \nu , % \simeq  1.249\ A_m(t) \Gamma_m(t), 
\\
    \nu_m(t_n) &=& \frac{\sum_{\nu\in W_m} \nu [P_m(\nu,t_n) - B_m(t_n)] }{\sum_{\nu\in W_m} [P_m(\nu,t_n) - B_m(t_n)]} .
\end{eqnarray}
In the above equations, 
the sums are computed over a frequency interval 
$W_m = [\overline{\nu}_m - 3\overline{\Gamma}_m/2 , \overline{\nu}_m+ 3\overline{\Gamma}_m/2]$, which is centered at the average mode frequency $\overline{\nu}_m$. 
The frequency resolution is denoted by  $\Delta \nu \approx 8$ nHz and the background power for the $n$-th time segment is denoted by $B_m(t_n)$.
The background power depends on the phase of the solar cycle and is obtained by computing the median power outside the peak of power. 
The time stamps for which the mode power is not significant are discarded.
 A sample power spectrum of the $m=8$ mode in the time window centered at $2010.5$ is shown in Fig.~\ref{fig:m8_spectra_tstamp}. 

\begin{figure*}[h]
    \centering
    \includegraphics[width=1\textwidth]{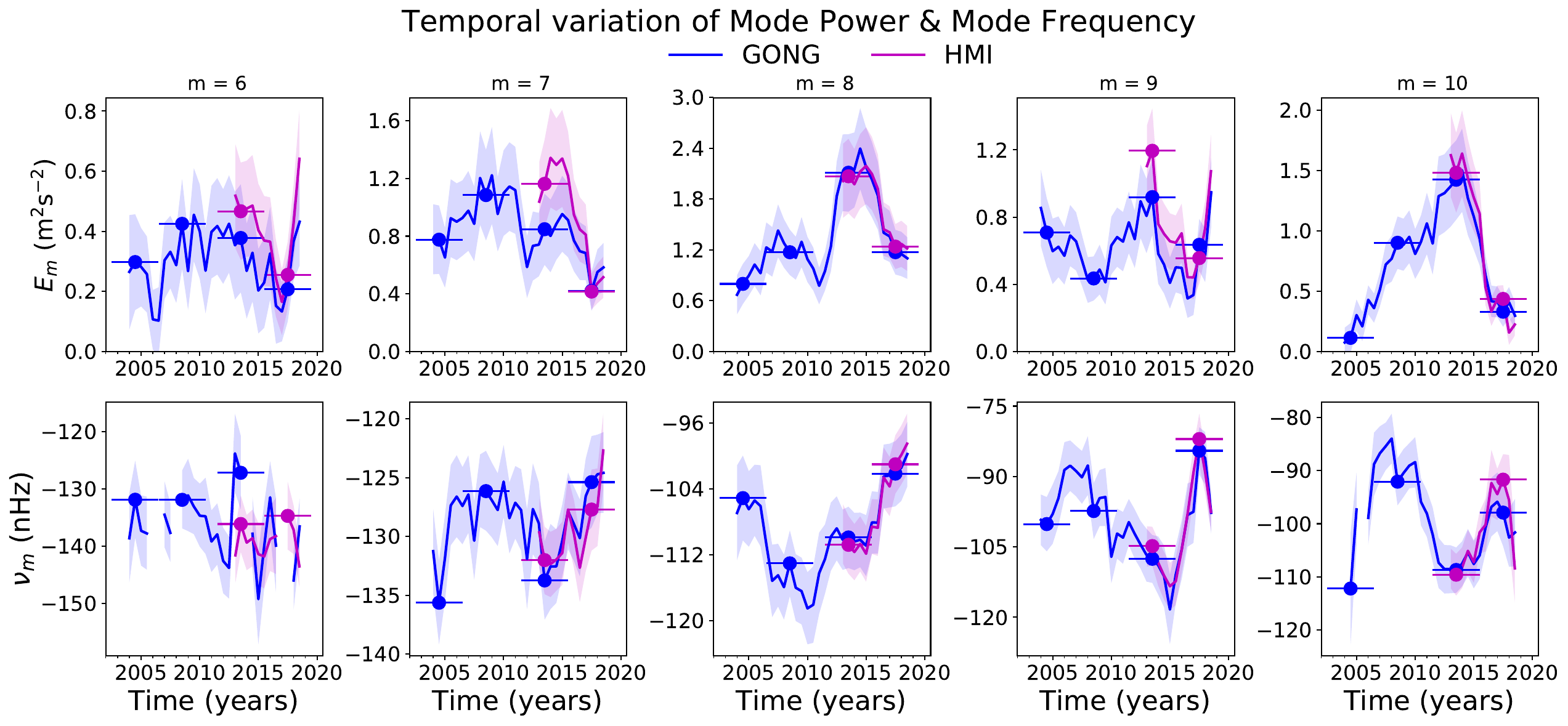}
    \caption{The temporal variation of mode power $E_m$ (top row) and mode frequency $\nu_m$ (bottom row) of the equatorial Rossby modes with $m$ from 6 to 10 from GONG and HMI data. The shaded regions represent the $68\%$ confidence bounds  estimated from Monte Carlo simulations.}
    \label{fig:totpow_freq}
\end{figure*}

The $E_m$ and $\nu_m$ of different modes as  functions of time are plotted in Fig.~\ref{fig:totpow_freq}. The shaded regions represent the $68\%$ confidence interval on each parameter, estimated from Monte Carlo simulations. 

\section*{Discussion}

\begin{figure*}[h]
    \centering
    \includegraphics[width=\textwidth, trim={0 0.3cm 0 1.5cm},clip]{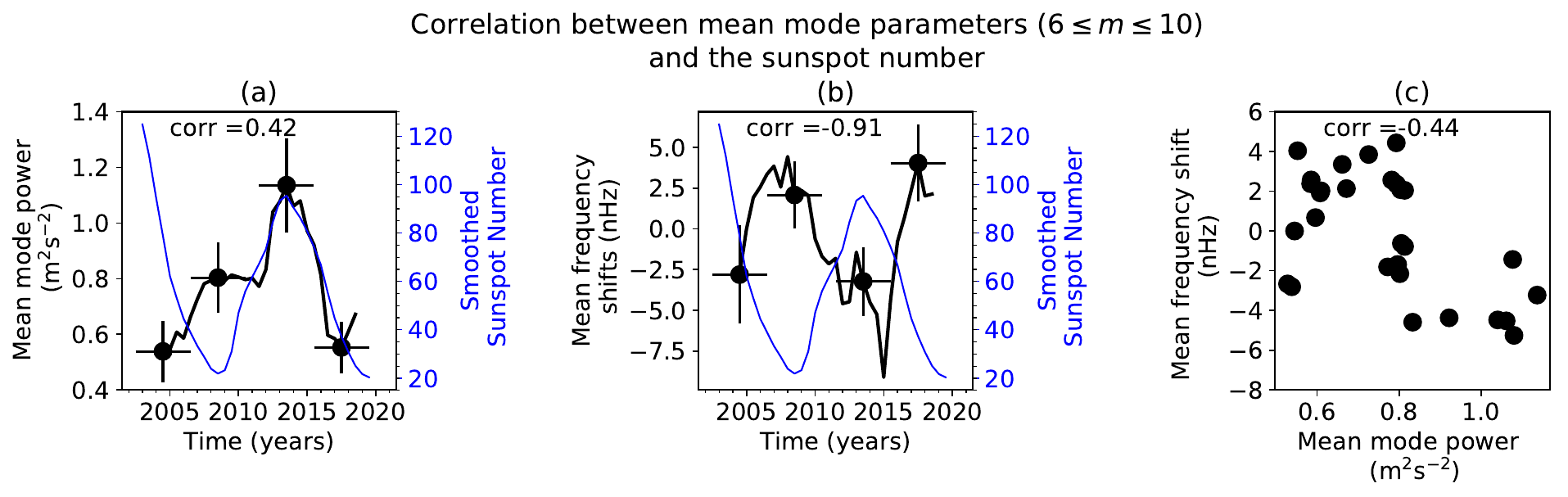}
    \caption{(a) Mode power $\langle E_m(t_n) \rangle$  and (b) mode frequency shift $\langle \nu_m(t_n) - \overline{\nu}_m \rangle$  averaged over $m$ from 6 to 10 as  functions of time.  The Pearson correlation coefficients between these and the smoothed sunspot number are given on the plots. Only the non-overlapping  time segments (points with error bars) were used in the computation of the correlation coefficients. 
    (c) Scatter plot of mode frequency shifts versus mode power. The Pearson correlation coefficient is $-0.44$.}
    \label{fig:correlation}
\end{figure*}

To study the relationship between the mode parameters and the solar cycle, we obtain  the sunspot numbers (SSN)  from the \cite{sidc}. 
In order to reduce the noise level at any given time stamp, we average the mode parameters over $m$ (6 to 10). These $m$-averaged parameters are plotted in Fig.~\ref{fig:correlation}. The Pearson correlation coefficients between the parameters and the SSN are computed using only  non-overlapping (statistically independent) time segments. We find a positive correlation coefficient ($0.42$) between the mode power and SSN, while the frequency shifts show a very strong anti-correlation ($-0.91$) with the SSN.
An anti-correlation ($-0.44$) is observed between the mode power and the frequency shift. 

Our observations align broadly with the HMI findings from \cite{Waidele2023}. 
Using the GONG data, we confirm that the anticorrelation between the frequency shifts and SSN is very strong through both cycles 23 and 24.
However, we find that the correlation obtained between the mode power and the SSN is not as strong  during cycle 23 as during cycle 24.  
We note that the observed frequency shifts have  a sign opposite to that given by a model \citep{Goddard2020} based on the perturbation of the modes by the torsional oscillations (time-varying rotation). Thus we infer that magnetic effects must play an important  role in  the solar-cycle variations  of the equatorial Rossby mode parameters.

\section*{Acknowledgements}
We acknowledge the support from ERC Synergy Grant WHOLE SUN 810218 and Deutsche Forschungsgemeinschaft (DFG, German Research Foundation) through SFB 1456/432680300 Mathematics of Experiment, project C04. 
Data were acquired by GONG instruments operated by NISP/NSO/AURA/NSF with contribution from NOAA. The HMI data are courtesy of NASA/SDO and the HMI Science Team.

\end{document}